**Guidelines for estimating causal effects in pragmatic randomized trials**

Eleanor J. Murray[1], Sonja A. Swanson[2], Miguel A. Hernán[1,3,4]

[1]Department of Epidemiology, Harvard T.H. Chan School of Public Health, Boston, MA, USA
[2]Department of Epidemiology, Erasmus Medical Center, Rotterdam, The Netherlands
[3]Department of Biostatistics, Harvard T.H. Chan School of Public Health, Boston, MA, USA
[4]Harvard-MIT Division of Health Sciences and Technology, Cambridge, USA.

**Corresponding author**:
Eleanor Murray
Department of Epidemiology
Boston University School of Public Health
322 Talbot East
Boston, MA, USA, 02118
Email: ejmurray@bu.edu

**Funding**: This study was supported through a Patient Centered Outcomes Research Institute (PCORI) award (ME-1503-8119). All statements in this report, including its findings and conclusions, are solely those of the authors and do not necessarily represent the views of the Patient-Centered Outcomes Research Institute (PCORI), its Board of Governors or Methodology Committee. The funder approved the protocol but had no role in the conduct, analysis, or reporting of study findings.

**Word count**: 9882

**Abstract**
Pragmatic randomized trials are designed to provide evidence for clinical decision-making rather than regulatory approval. Common features of these trials include the inclusion of heterogeneous or diverse patient populations in a wide range of care settings, the use of active treatment strategies as comparators, unblinded treatment assignment, and the study of long-term, clinically relevant outcomes. These features can greatly increase the usefulness of the trial results for patients, clinicians, and other stakeholders. However, these features also introduce an increased risk of non-adherence, which reduces the value of the intention-to-treat effect as a patient-centered measure of causal effect. In these settings, the per-protocol effect provides useful complementary information for decision making. Unfortunately, there is little guidance for valid estimation of the per-protocol effect. Here, we present our full guidelines for analyses of pragmatic trials that will result in more informative causal inferences for both the intention-to-treat effect and the per-protocol effect.

# 1. Introduction

Pragmatic randomized trials are key tools for research on the comparative effectiveness of medical interventions. Like other randomized trials, pragmatic trials are designed to compare strategies for the prevention, diagnosis, and treatment of diseases. The CONSORT extension for pragmatic trials defines these trials as "designed to measure effectiveness; that is whether an intervention works when used in usual conditions of care" (1,2) Unlike other randomized trials which are designed to seek regulatory approval, pragmatic trials are specifically designed to address real-world questions about options for care and therefore to guide decisions by patients, clinicians and other stakeholders. Therefore, characteristics of a pragmatic trial include heterogeneous or diverse patients and care settings, clinically relevant comparators (e.g., usual care rather than placebo), unconcealed assignment to treatment, and follow-up time long enough to study long-term clinical outcomes without having to rely on surrogates (3,4).

While pragmatic trials are useful to guide decision making, they are also especially vulnerable to post-randomization confounding from incomplete adherence and post-randomization selection bias from loss to follow-up (5). That is, pragmatic trials are especially subject to many of the biases that we have learned to associate exclusively with observational follow-up studies. Yet no standardized methodology for causal inference from pragmatic trials has been proposed, with most investigators exclusively relying on the intention-to-treat principle.

The emphasis on the intention-to-treat principle is the historical consequence of many early randomized trials being short, small, double-blinded, tightly controlled experiments among highly selected patients who largely adhered to the assigned treatment and who were rarely lost to follow-up. Because these experiments are designed to minimize post-randomization confounding and selection bias, intention-to-treat analyses provide reasonable measures of the treatment effect. However, the almost exclusive reliance on the intention-to-treat principle in pragmatic trials is worrisome because, under real-world conditions, the effect estimates may be profoundly impacted by non-adherence and loss to follow-up. As a result, pragmatic trials exclusively analyzed under the intention-to-treat principle may not be quantifying the causal effect of primary interest for patients, clinicians, researchers, and other stakeholders (6).

Yet there is little methodological guidance for pragmatic trials beyond intention-to-treat. For example, the Biostatistics and Study Design Core of the National Institutes of Health Collaboratory has provided insightful guidance about many technical aspects of pragmatic trials (7), but it presupposes an intention-to-treat analysis. Similarly, the Clinical Trials Transformation Initiative—a public-private partnership comprising government agencies, industry, patient advocacy groups, professional societies, investigator groups, academic institutions, and other interested parties—promotes helpful proposals to improve clinical trials, but does not directly address causal inference standards (8). Emphatic calls to overcome the barriers for the conduct of large, simple pragmatic trials also presuppose an intention-to-treat analysis (9).

Here we propose causal inference guidelines for the analysis of pragmatic clinical trials. Because some of our recommended analyses require data on post-randomization treatment decisions and covariates, embracing these guidelines will require a revised framework for the design of pragmatic trials and other trials with substantial loss to follow-up or non-adherence. In fact, these guidelines are relevant for trials with individual-level randomization and parallel groups, even if they are not defined as pragmatic (conversely, some of the case studies described below are based on non-pragmatic trials but apply equally to pragmatic ones). Therefore, these guidelines complement recent efforts to go beyond the intention-to-treat principle in regulatory trials (10). Crossover trials and cluster randomized trials may also benefit from a consideration of these guidelines, but these trials have special features which are beyond the scope of this paper. Also, this document is primarily concerned with issues related to bias in the effect estimates of a pragmatic trial, rather than with generalizability and transportability of those effect estimates to other populations.

This document is organized as follows. The next section discusses two options for causal effects that can be estimated from pragmatic trials: the intention-to-treat effect and the per-protocol effect. Section 3 emphasizes the need to measure effects on an absolute, rather than relative, scale. Sections 4, 5, and 6 provide guidelines for the estimation of the intention-to-treat effect, the per-protocol effect for point interventions, and the per-protocol effect for treatment strategies sustained over time. Throughout, we feature case studies to emphasize key points.

**2. Choice of causal effect: Intention-to-treat effect vs. Per-protocol effect**

The first step towards a relevant causal analysis of pragmatic trials is a precise definition of the causal effect of interest (also known as the causal estimand). An explicit specification of the causal effect is important to engage (i) patients and stakeholders in a conversation about what they expect to learn from the trial, and (ii) investigators in a conversation about design and data analysis choices. A clear distinction should be made between the causal effect of interest and the method by which an estimate of that effect is obtained. Here, we will consider two causal effects:

- *Intention-to-treat effect*: the effect of being assigned to the treatment strategies, regardless of treatment actually received during the follow-up. In trials where initiation of the strategies co-occurs with randomization, the intention-to-treat effect is the effect of initiation of the treatment strategies, regardless of subsequent adherence to them.
- *Per-protocol effect*: the effect of receiving the assigned treatment strategies throughout the follow-up as specified in the study protocol.

Some authors have referred to the intention-to-treat effect as the "treatment policy effect" (10). This is an unfortunate term because, although the intention-to-treat effect is indeed a contrast of treatment policies or strategies (specifically the policies "assign to a treatment strategy, then do whatever you want"), so is the per-protocol effect. We now review the relative advantages and disadvantages of these two causal effects.

*2.1 The intention-to-treat effect*

The intention-to-treat effect, which is the default target of many randomized trials is appealing for several reasons, included the perceived simplicity with which a valid estimate can be obtained. However, the strengths of the intention-to-treat effect come at the cost of limited utility for decision-making, interpretability, and external validity.

The primary argument in favor of the intention-to-treat effect is, of course, that assignment is randomized. Therefore, we expect that a simple analysis that estimates the association between assignment and outcome—an intention-to-treat analysis—will yield an unconfounded estimate of the intention-to-treat effect. A second argument is that an intention-to-treat analysis provides a valid statistical test of the null hypothesis of no treatment effect in blinded trials. That is, if the treatment effect is null then the intention-to-treat effect is expected to be null regardless of the actual adherence pattern in the trial. This null preservation property is a desirable property. The intention-to-treat effect is said to be conservative ("biased towards the null") in placebo-controlled trials because the magnitude of the intention-to-treat effect is somewhere in between the null value and the true effect of treatment. Conservativeness is often presented as a desirable property of the intention-to-treat effect.

However, these arguments are less compelling for pragmatic trials. First, pragmatic trials are not double-blind randomized trials. Therefore, the effect of being assigned to treatment—the intention-to-treat effect—will typically be a combination of the effect of the treatment under study and of any other patient and physician's behavioral changes triggered by the assignment itself. Quantifying this combination effect may be of practical interest in some settings, but it follows that intention-to-treat analyses may not preserve the null: the intention-to-treat effect estimate may not be null, even if the effect of treatment is null, because the intention-to-treat incorporates effects of assignment not mediated by treatment itself. Second, pragmatic trials are rarely placebo-controlled trials. Therefore, the intention-to-treat may not be conservative, even when the treatment effects are monotonic (11), because pragmatic trials usually compare treatment strategies with potentially differential adherence. Regardless, conservativeness is an undesirable property in trials studying harms because a conservative treatment effect can result in missed harms, and in non-inferiority trials because a conservative treatment effects can also lead to erroneous claims of non-inferiority (12,13).

Intention-to-treat estimates are also difficult for patients, clinicians, and other decision-makers to interpret (6) because the intention-to-treat effect is agnostic to any treatment decisions made after baseline, including discontinuation or initiation of the treatment strategies of interest, use of concomitant therapies, or any other deviations from protocol. As a result, the magnitude of the intention-to-treat from a given trial depends on the particular patterns of deviation from protocol that occur during the conduct of the trial. Two trials of the same treatment strategies, conducted in the same population, could have different intention-to-treat effects if adherence patterns differed, and both would be internally valid effects of assignment to treatment. The external validity of the intention-to-treat effect can be poor and difficult to assess because patterns of adherence in the trial may not reflect those outside the trial. In fact, the publication of the trial results may lead to alterations in adherence patterns outside the trial (14).

Further, the intention-to-treat effect is often considered to estimate treatment effectiveness (that is, the effect of treatment under realistic conditions or in everyday practice) rather than efficacy (that is, the effect of treatment under ideal conditions including perfect adherence). However, this distinction between efficacy and effectiveness is artificial (in fact, we don't distinguish between "safety" and "safetiness" (5)) and of unclear relevance for decision makers. For example, a patient deciding whether to take cholesterol lowering medication is usually interested in its effectiveness, rather than the effectiveness estimated from a study in which many trial participants did not adhere to a correct use (6,12). In general, patients are mostly interested in the effect of treatment in everyday practice when taken as instructed, an effect that corresponds to the common definition of neither "efficacy" nor "effectiveness". We therefore largely avoid the use of those terms here.

Finally, it is not universally true that traditional intention-to-treat analyses provide a valid estimate of the intention-to-treat effect. In trials with incomplete follow-up of participants, losses to follow-up may introduce selection bias (5,15). Eliminating this bias requires adjusting for pre- and post-randomization predictors of loss to follow-up that are also prognostic factors.

To summarize, the intention-to-treat effect from a given trial measures the effect of treatment assignment, not of treatment itself, under whatever level of adherence to protocol took place in that trial. Therefore, intention-to-treat estimates, even if appropriately adjusted for loss to follow-up, provide incomplete information for patients who are interested in the benefit-risk profile of the treatment when taken as instructed.

*2.2 The per-protocol effect*

The per-protocol effect overcomes many limitations of the intention-to-treat effect. The per-protocol effect is closer than the intention-to-treat effect to what patients are mostly interested in learning from pragmatic trials (Case Study A), and is often the implicit target of inference for investigators too.

When investigators say that the intention-to-treat is "biased towards the null", the implication is that the intention-to-treat effect is a biased estimate of the per-protocol effect (a bias sometimes referred to as "performance bias" (16,17)). Thus, the investigators' language indicates that the investigators are really interested in comparing the interventions implemented during the follow-up as specified in the protocol (i.e., the per-protocol effect) and not just in the effect of assignment to the interventions at baseline (i.e., the intention-to-treat effect).

An added advantage of the per-protocol effect is that its interpretation does not depend on a trial-specific degree of adherence (14), which makes it a potentially more transportable effect. For example, in a cancer prevention trial, participants were randomized to either an invitation to receive a screening colonoscopy or to usual care (with no colonoscopy screening) (18). The intention-to-treat effect for this trial quantifies the effect of being invited to a screening colonoscopy in a population in which about 30% of individuals declined the invitation. Because the proportion and type of people who reject a colonoscopy varies across populations, the intention-to-treat effect from this trial is difficult to transport to other populations. In contrast, the per-protocol effect would quantify the effect of receiving the colonoscopy under perfect adherence in the studied population, which may be closer to the effect under perfect adherence in similar populations.

| *Case Study A: Intention-to-treat vs. per-protocol effect* |
|---|
| The Women's Health Initiative (WHI) conducted a randomized trial to estimate the health effects of postmenopausal estrogen plus progestin hormone therapy versus placebo (19,20). Analyses to estimate both the intention-to-treat effect and the per-protocol effect were conducted (the latter (21) made the simplifying, but plausible, assumption that the number of women with contraindications for hormone therapy during the follow-up was negligible.) The estimated intention-to-treat hazard ratio of breast cancer was 1.25 (95% CI: 1.01, 1.54). This 25% increased risk reflects both the actual effect of hormone therapy and the incomplete adherence to the assigned treatment: only 58% of women in the hormone therapy arm, and 62% in the placebo arm, were still taking their assigned treatment at 6 years.<br>In contrast, the estimated per-protocol hazard ratio of breast cancer was 1.68 (95% CI: 1.24, 2.28). This effect estimate suggests that continued use of postmenopausal estrogen plus progestin hormone use increases the risk of breast cancer by 68% rather than 25%. Women presented with the intention-to-treat effect estimate only might reasonably argue that they did not receive full information about their increased risk of breast cancer if they took treatment as instructed. |

Yet, despite its relevance for decision making by patients and clinicians, the per-protocol effect has been historically eschewed in randomized trials. A fundamental problem is that valid estimation of the per-protocol effect cannot be guaranteed by randomization because unbiased estimation generally requires adjustment for prognostic factors that predict adherence. Faced with the choice between a less relevant effect that is expected to be unbiased in the absence of substantial loss to follow-up (the intention-to-treat effect) and a more relevant effect that may be biased (the per-protocol effect), researchers chose the former.

A consequence of the general skepticism about the per-protocol effect is that little emphasis has been given to studying the conditions under which it can be correctly estimated. This in turn has led to the design and conduct of randomized trials that are intrinsically ill-equipped to quantify the per-protocol effect, which often results in the intention-to-treat effect being the only viable option. To break this vicious circle, one needs to realize that unbiased estimation of the per-protocol effect in a randomized trial generally requires 3 elements:

1) Unambiguous specification of the treatment strategies in the trial protocol. A common misconception is that trials compare treatments; rather, they compare treatment strategies (13,14). Individuals assigned to a particular treatment must be allowed to discontinue, modify, or switch the treatment when clinically indicated (e.g., due to toxicity or lack of effectiveness) and still remain on protocol, both for ethical reasons and in order to ensure the protocol is of interest given current knowledge and medical practice (14). Therefore, adherence to the protocol is not defined as "taking treatment continuously no matter what" but as, for example, "taking treatment until serious side effects arise" (Case Study B).
2) Collection of data on prognostic factors that predict adherence to the protocol. Most pragmatic trials compare treatment strategies that are sustained over the follow-up. As a result, adherence to the strategies may be affected by both pre- and post-randomization factors. When the treatment strategy is a point intervention, prognostic factors that affect adherence may not be required for effect estimation but are useful for characterizing the compliers (see Section 5.2).
3) Adjustment for prognostic factors that predict adherence to the protocol. When the value of the post-randomization factors is affected by prior adherence, the adjustment needs to be carried out via g-methods (11), as discussed below.

To summarize, valid estimation of the per-protocol effect generally requires a specification of the treatment strategies, adequate data collection, and appropriate adjustment. In addition, at the design stage, researchers may want to ensure a sufficient sample size to estimate the per-protocol effect with sufficient precision. Given the complexities involved in per-protocol effect estimation (see below), simulations of the data generation process will often be necessary for approximate sample size calculations.

| *Case study B: Specification of the treatment strategies in the protocol* |
|---|
| The Candesartan in Heart Failure: Morbidity and Mortality (CHARM) trial was a secondary prevention trial aimed at reducing mortality and cardiac-related hospitalization among individuals with symptomatic congestive heart failure (22). The estimated intention-to-treat mortality hazard ratio for candesartan vs. placebo was 0.89 (95% CI: 0.82, 0.97) (22). To quantify the extent to which this intention-to-treat effect may underestimate the effect of treatment, we could estimate the per-protocol effect. <br><br> But the per-protocol effect cannot be estimated unless it is explicitly defined. Importantly, the per-protocol effect is *not* the effect of taking candesartan continuously but rather the effect of adhering continuously to the treatment strategy specified in the protocol. In CHARM, the per-protocol effect is the effect of taking candesartan continuously until the occurrence of abnormal renal function or hypotension. Individuals who discontinued their assigned treatment after experiencing either of these conditions need to be considered adherent to the protocol for the remainder of the trial, despite no longer taking their assigned medication (23). |

> The trial protocol also specified that treating clinicians could approve the discontinuation or dosage reduction of an individual's assigned treatment at their discretion based on a perception of medication intolerance (22). Therefore, one can define an alternate per-protocol effect in which patients are considered to be adherent upon clinician-approved treatment discontinuation (23).

**Guideline:**
  I.  **To adequately guide decision making by all stakeholders, report estimates of both the intention-to-treat effect and the per-protocol effect, as well as methods and key conditions underlying the estimation procedures.**

### 3. Assessment of effect magnitude and heterogeneity

Many pragmatic trials compare the risk of developing a health outcome, as opposed to non-binary outcomes like blood pressure or quality of life, under different treatment strategies. In these trials, the effects are often reported as relative risks, or related measures such as hazard ratios. In addition, most trials also report absolute risks, or related measures such as cumulative incidence curves or survival curves, but reporting absolute difference measures is less common (6).

Relative risks provide information on the direction of the causal effect, but not on the absolute benefit (or harm) of the treatment. For example, a relative risk of 3 for treatment versus no treatment indicates a 3-fold increased risk under treatment compared with no treatment, but carries no information on the absolute increase in risk. If the absolute risk under no treatment were 10%, then a relative risk of 3 would translate into an increased risk of 20 percentage points. In contrast, if the absolute risk under no treatment were 1%, then a relative risk of 3 would translate into an increased risk of only 2 percentage points. Since the primary purpose of pragmatic trials is to guide decision-making for patients and clinicians, absolute risks and their differences on the additive scale should always be presented (Case Study C). Patients prefer absolute measures of occurrence when participating in shared decision-making (6,24).

> *Case Study C: Absolute vs. relative risks*
>
> Case Study #1 described the intention-to-treat and per-protocol estimates of the effect of postmenopausal hormone therapy on breast cancer. We concluded that women considering the initiation of therapy would be misinformed if provided with the intention-to-treat effect (25% increased risk) but not the per-protocol effect (65% increased risk). However, our reasoning was based on relative risk estimates only.
>
> Had we used absolute risks, our conclusion would not have been as strong. The estimated intention-to-treat difference in 8-year risk of breast cancer incidence if all women had been assigned to hormone therapy versus if all had been assigned to placebo was only 0.83 percentage points (95% CI: -0.03, 1.69), while the estimated per-protocol difference in 8-year risk if all women had continuously used hormone therapy versus all women had used no hormone therapy was 1.44 percentage points (95% CI: 0.52, 2.37) (21). When providing effect estimates on the additive scale, both the intention-to-treat and per-protocol estimates suggest that the 8-year risk increases by less than 1.5 percentage points.

The above discussion is especially relevant when a goal of the trial is to quantify whether there is treatment effect heterogeneity, that is, whether the effect varies across subgroups of the study population defined by their baseline characteristics, e.g., men and women, younger and older, diabetics and nondiabetics. Estimating treatment effects in different strata of trial participants is a natural goal of pragmatic trials, since this can improve decision-making. In fact, patients report a high degree of interest in stratum-specific treatment effects that provide them with some guidance on how a medical intervention may work for individuals similar to themselves (6).

When a goal is the identification of effect heterogeneity, the trial protocol will typically specify the participants' strata within which the effects will be estimated. Ideally, clinicians, patients, and other stakeholders will be consulted to pre-specify the subgroups of interest. Then the effect estimates will be obtained separately in each of the subgroups of interest (Case Study D). New methods that allow estimation of subgroup-specific causal effects without requiring pre-specification of the strata of interest are also being developed (25,26). Importantly, when the goal is to identify the patient subgroups that are more likely to benefit (or be harmed) by treatment, the effect estimates must be measured on the additive, not the multiplicative, scale (27). For example, the comparison of risk differences, not risk ratios, across subgroups is the correct procedure in this setting.

A final comment. though most trials already present comparison of absolute risks to quantify the intention-to-treat effect, it is important to bear in mind these considerations also when estimating per-protocol effects. The valid estimation of per-protocol effects typically requires adjustment for prognostic factors, and some researchers seem to be under the impression that adjusted analyses can only provide relative risk measures. On the contrary, as discussed below, adjusted analyses can yield adequately standardized absolute risks and risk differences. The choice of the scale (additive or multiplicative) on which the effect is measured does not dictate the analytic approach (28).

| *Case Study D: Heterogeneity of treatment effects* |
| --- |
| In the Women's Health Initiative (WHI) estrogen plus progestin trial (Case Study #1), the intention-to-treat analysis was conducted in 23 pre-specified subgroups (29) defined by demographic and clinical characteristics for which heterogeneity of treatment effects was expected.<br><br>For example, the intention-to-treat odds ratio of coronary heart disease for hormone therapy versus placebo ranged from 0.76 among women with low cholesterol at baseline to 2.03 among women with high cholesterol at baseline. Unfortunately, absolute risks or rates were not presented for these sub-groups so heterogeneity could not be evaluated on the additive scale. |

**Guideline:**
    **II.    Report absolute risks and their differences, as well as their ratios, for discrete outcomes.**
    **III.   Heterogeneity of treatment effects can be reported using subgroup analyses that use the additive scale to measure the effect of interest. Patients and advocates should be included in *a priori* specification of subgroups.**

**4. Estimation of the intention-to-treat effect**
We refer to the analyses aimed at estimating the intention-to-treat effect as intention-to-treat analyses. In a large pragmatic trial with complete follow-up and no competing events for the outcome, the intention-to-treat analysis is straightforward: compare the observed outcome distribution between trial arms. For example, in a trial of aspirin and mortality, the intention-to-treat risk ratio at 5 years is the ratio of the observed 5-year mortality risk in those assigned to daily aspirin divided by the 5-year mortality risk in those assigned to no aspirin. That is, in this setting the intention-to-treat effect can be validly estimated without adjustment for covariates. We now review three settings in which analyses to estimate the intention-to-treat effect require additional measures—imbalanced prognostic factors at baseline, competing events, and losses to follow-up—and provide guidelines. These guidelines also apply to the estimation of per-protocol effects (which is discussed in Sections 5 and 6).

*4.1 Imbalanced prognostic factors at baseline*

Randomization cannot ensure that the distribution of all prognostic factors is identical between the trial arms. For example, in an aspirin trial the proportion of smokers may be, just by chance, greater in the group assigned to aspirin than in the group assigned to no aspirin. As a result of these chance imbalances, the effect estimate may be far from the true effect. Some authors refer to the difference between a study-specific effect estimate and the true effect as "random confounding" because, in a particular study, the problem is indistinguishable from the systematic confounding that results from systematic imbalances of prognostic factors (30). Practically speaking, it is advisable to deal with any imbalances in intention-to-treat analyses, regardless of whether they are random or systematic, via adjustment for the imbalanced prognostic factors (31). Of course, in well conducted randomized trials, only random and not systematic imbalances are expected.

How to best choose the prognostic factors that need to be adjusted for in intention-to-treat analyses remains debated. While there is consensus that the use of significance testing is strongly discouraged (32), several other options exist. One option is to pre-specify important prognostic factors for adjustment (Case Study E1). Another possibility is to pre-specify a maximum acceptable imbalance which, if exceeded, will trigger adjustment for that factor. Pre-specification of a maximum acceptable imbalance could be based on discussion with subject matter experts about the degree of confounding likely in an observational study, or estimated using an e-value calculation based on assumptions about likely unadjusted intention-to-treat estimates (33).

Yet neither option provides guidance about what to do when a large imbalance is found for a strong prognostic factor that was not pre-specified. In this case, the recommended course of action is to adjust for that factor (31,34), even if only as a sensitivity analysis (Case Studies E2, E3). However, such adjustment raises concerns about the validity of the statistical inferences. A principled approach to adjustment that preserves the inferential properties of the statistical analysis would need to be based on doubly robust procedures (35–37). However, software for doubly robust methods is not yet readily applicable to all the settings (e.g., survival analysis with time-varying treatments and covariates) that investigators may encounter when designing and analyzing pragmatic trials.

To preserve the marginal (unconditional) interpretation of intention-to-treat effect estimates in the study population, adjustment for baseline covariates is better carried out via standardization, inverse probability weighting (Case Study E1) or, preferably, doubly-robust methods (11,38,39). Adjustment methods like outcome regression or propensity score analysis may result in more precise intention-to-treat effect estimates (40–42) but the validity of these estimates rely on the assumption of effect homogeneity across levels of the covariates.

> *Case study E1: Adjusting for pre-specified baseline prognostic factors*

In the Candesartan in Heart Failure: Morbidity and Mortality (CHARM) trial, a list of important baseline prognostic factors for all-cause mortality among individuals with heart failure was pre-specified, including heart disease risk factors, medical history, and use of medical treatments. Table 1 described the distribution of these prognostic factors separately in the candesartan and placebo arms. The unadjusted intention-to-treat estimate of the effect of randomization to candesartan versus placebo on all-cause mortality was a hazard ratio of 0.91 (95% CI: 0.83, 1.00) (22). After adjustment for all pre-specified baseline prognostic factors, the estimated effect of randomization to candesartan versus placebo on all-cause mortality was a hazard ratio of 0.90 (95% CI: 0.82, 0.99) (22).

In a secondary analysis of the CHARM data, we also estimated the intention-to-treat effect standardized across these baseline prognostic factors to improve interpretation – standardization allows the interpretation as an average effect among the trial population, rather than as a conditional effect "holding covariates constant". The average intention-to-treat effect of randomization to candesartan versus placebo was 0.89 (95% CI: 0.82, 0.97) (23). The absence of material differences across of all these estimates is expected in randomized trials which, like CHARM, have small imbalances in baseline covariates.

*Case study E2: Adjusting for imbalanced baseline prognostic factors*

The Norwegian Colorectal Cancer Prevention (NORCCAP) trial was a primary prevention trial for colorectal cancer incidence and mortality comparing a one-time screening sigmoidoscopy with no screening(18). In the original trial protocol, all individuals aged 55 to 64 years old living in two regions of Norway in 1998 were eligible for participation in the NORCCAP trial. A random sample of these individuals were sent an invitation for a screening sigmoidoscopy. However, in 2000, the funding agencies decided to increase the eligibility to include individuals aged 50 to 54 years old. The same random selection process was used for this new age group, but because of the larger size of this post-war birth cohort, the ratio of invitation group to control group participants was lower among younger individuals. Since age is a strong prognostic factor for colorectal cancer, this design change artificially induced confounding by age—which had not been pre-specified for adjustment in the original trial design. However, since there was a substantial imbalance in age between trial arms, the primary analysis reported age-standardized effect measures. For example, the age-standardized intention-to-treat difference for the effect of randomization to screening invitation versus randomization to no invitation was a 28.4 percentage point reduction in the rate of colorectal cancer diagnoses over the 10 years of follow-up.

*Case study E3: Adjusting for pre-specified and imbalanced baseline prognostic factors*

In the Women's Health Initiative (WHI) estrogen plus progestin trial (Case Study #1), intention-to-treat analyses were adjusted for a set of pre-specified baseline covariates: age, evidence of coronary heart disease at enrollment, and randomization status in a low-fat diet sub-trial (43).

The intention-to-treat analysis for coronary heart disease was further adjusted for history of coronary artery bypass graft (CABG) or percutaneous transluminal coronary angioplasty (PTCA) at enrollment (29). These covariates, which were not pre-specified, were slightly imbalanced between randomization arms: the prevalence of CABG/PTCA was 1.1% in the estrogen plus progesterone arm was 1.1% compared with 1.5% in the placebo arm. This variable was chosen because the 0.4% difference was "statistically significant" (p=0.04) whereas other variables with greater differences were not adjusted for (43). Adjustment did not materially change the intention-to-treat estimates.

**Guideline:**
IV. **Pre-specify important prognostic factors for the outcome and the maximum acceptable difference in the distribution of these factors between groups. When one or more prognostic factor meets the threshold for imbalance, adjust via standardization, inverse probability weighting or, preferably, doubly-robust methods.**
V. **In sensitivity analyses, adjust for large imbalances in any important prognostic factors, regardless of whether they have been pre-specified.**

*4.2 Competing events*

A competing event is an event that precludes the outcome from happening (44). For example, if we are interested in studying whether daily aspirin use decreases the risk of stroke, then death from causes other than stroke is a competing event because individuals who die are no longer at risk for stroke. In these settings, informed decision analysis requires information about both the risk of the competing event by treatment group and the risk of the event interest among those who survived the competing events.

The presence of competing events implies that the definition of intention-to-treat effect provided in Section 2 is incomplete. In order to decide how to account for competing events in intention-to-treat analyses, we first need to choose among several possible definitions of intention-to-treat effect (i.e., causal estimands) (11,44). For example, in our aspirin example, we consider any of the following definitions of intention-to-treat effect:

1) The total effect of assignment to aspirin versus no aspirin on stroke in a population with the same death rate as the trial population
2) The direct effect of assignment to aspirin versus no aspirin on stroke if no one had died before having a stroke;
3) The total effect of assignment to aspirin versus no aspirin on stroke among those individuals who would never have died before having a stroke during the trial regardless of which assignment they had been given
4) The total effect of assignment to aspirin versus no aspirin on death or stroke, whichever happens first

The choice among these (or other (45)) effects implies trade-offs between ease of estimation, tenability of the assumptions, and interpretability of the effect estimate (Case study F). The first effect, which is estimated by a simple contrast of the cumulative incidence in each group, does not allow us to distinguish whether the effect on the outcome of interest is mediated through the effect on death (e.g., the more people die in a treatment group, the fewer people can develop stroke). The second effect, which is estimated by censoring individuals if/when they develop the competing event and adjusting for potential selection bias, measures the direct effect on the outcome that is not mediated by the competing event. This effect, however, is hard to interpret as no well-defined intervention exists to prevent all deaths from causes other than the event of interest. The third effect, which requires strong assumptions, is also hard to of limited relevance to clinical decision-making because it is restricted to an unidentifiable subset of the population (known as a principal stratum). Finally, the fourth effect eliminates the competing events problem by estimating the risk of the combined outcome, but can fundamentally change the research question of interest (11,44).

Given that none of the available approaches to deal with competing events is clearly superior to the others, it may be advisable to choose the simplest estimand (the total effect on the event of interest in the entire study population; the first effect listed above) as the target of the primary analysis and to complement it with sensitivity analyses that estimate the direct effect after confirming that treatment assignment has an effect on the competing event. Importantly, the choice among several causal estimands is generally only possible in survival (e.g., failure time) analyses. When the outcome is a variable measured at the end of follow-up (e.g., blood pressure), then the only options are the direct effect estimated after censoring and the effect in those who would always have survived (46).

> *Case study F: The intention-to-treat effect in the presence of competing events*
>
> The Assessment of Multiple Intrauterine Gestations from Ovarian Stimulation (AMIGOS) trial compared maternal use of gonadotropins versus clomiphene during fertility treatments for the prevention of neonatal complications. In this trial, the outcome is only defined among pregnancies which ended in a live birth, that is, "no live birth" is a competing event for the event of interest (47).
>
> We can consider several causal estimands for the intention-to-treat effect in this trial (48), including:
> - the total effect of gonadotropins versus clomiphene on neonatal complications assuming that no complications can occur without live birth
> - the total effect of gonadotropins versus clomiphene on the composite outcome of either neonatal complications or no live birth
> - the direct effect of gonadotropins versus clomiphene on neonatal complications under the assumption that censoring due to failure to produce a live birth can be abolished and that all shared causes of the competing events and the event of interest can be adjusted for
> - the total effect of gonadotropins versus clomiphene on neonatal complications in pregnancies which would always result in a live births regardless of whether they were assigned to gonadotropins or clomiphene (an unidentifiable subset of the study population) under several untestable assumptions
>
> Each causal estimand will result in a different estimate of the intention-to-treat effect. Note that the last two estimands require strong assumptions that are not guaranteed to hold in any randomized trial.

**Guidelines:**
- **VI. In survival analyses with competing events, report both the risk of the competing event by treatment group and the risk of the event of interest among those who survived the competing event by treatment group.**
- **VII. In survival analyses with competing events, specify the intention-to-treat effect as the total effect of treatment assignment on the outcome of interest (the simplest analysis), and justify interest in any additional effects that are estimated.**

*4.3 Loss to follow-up*

Intention-to-treat analyses require that the outcomes of all trial participants are ascertained. If some participants are lost to follow-up (e.g., because they drop out of the study) and therefore their outcomes are unknown, a true intention-to-treat analysis is not possible.

In the presence of losses to follow-up, a common strategy is to conduct a complete case analysis in which participants are censored if/when they are lost to follow-up (or, for non-failure time outcomes, in which a participant's last available outcome measurement is carried forward). These pseudo-intention-to-treat analyses may introduce selection bias in the effect estimates (49). For example, if individuals are more likely to be lost to follow-up in the aspirin group and those who are lost to follow-up tend to have a worse health status, then a pseudo-intention-to-treat analysis will find better outcomes in the aspirin group compared with the no aspirin group, even if aspirin had no effect on the outcome.

Pseudo-intention to treat analyses provide valid estimates of the intention-to-treat effect if censoring due to loss to follow-up is completely at random with respect to the outcome (50). However, this assumption of non-informative censoring is untenable when, as it is common, loss to follow-up depends on pre- and post-randomization prognostic factors. Rather, one can make the weaker assumption that censoring is non-informative conditional on pre- and post-randomization covariates, and make sure that the pseudo-intention-to-treat analysis adjusts for those pre- and post-randomization prognostic factors (15,51). Adjusted pseudo-intention-to-treat analyses provide valid estimates of the intention-to-treat effect if censoring due to loss to follow-up is not informative within levels of the measured prognostic factors.

Because post-randomization prognostic factors may be affected by treatment assignment, adjustment methods that appropriately account for time-varying covariates, such as inverse probability weighting or the parametric g-formula, are usually preferable (5). Inverse probability weighting is typically easier to implement (Case Study G). Note that the implementation of these methods requires that post-randomization factors are measured regularly over follow-up to ensure that data will be available close to the time of censoring and that the model for censoring is correctly specified.

---

*Case study G: Adjusting for loss to follow-up when estimating the intention-to-treat effect*

In an open-label, multicenter randomized trial of antipsychotic medications to reduce self-reported symptoms among individuals with schizophrenia (52), the primary outcome of interest was Brief Psychiatric Rating Scale (BPRS) score at the end of follow-up (12 months). 450 individuals were assigned to one of two atypical antipsychotic treatments (combined here for simplicity), and 214 to conventional antipsychotics. However, only 430 individuals assigned to atypical treatments and 204 assigned to conventional completed at least one follow-up visit, and only 235 atypical arm and 130 conventional arm participants completed all follow-up visits (50). This high level of loss to follow-up could result in a biased intention-to-treat effect estimate.

A pseudo-intention-to-treat analysis yielded a difference in BPRS score of 0.42 units (95% CI: -2.36, 3.19) (50). However, loss to follow-up may have been related to pre- and post-randomization covariates. Adjusting for these via inverse probability weighting yielded an estimated intention-to-treat difference in BPRS score of -0.86 units (95% CI: -3.88, 2.15). Note that, as expected, the adjusted estimate is more imprecise than the unadjusted one.

---

**Guidelines:**
**VIII.     Ensure that the trial protocol specifies the collection of post-randomization time-varying prognostic factors that predict loss to follow-up, and appropriately adjust for these factors to reduce selection bias.**

*4.4 External validity*

An important goal of pragmatic randomized trials is to provide relevant and generalizable information to guide clinical decision-making. As such, the inclusion and exclusion criteria of pragmatic trials is often more broad than those of traditional randomized controlled trials. In order to ensure that pragmatic trial results are generalizable to the target population of interest, this population should be clearly and explicitly defined before the inclusion and exclusion criteria are defined.

In some cases, it will also be of interest to estimate what the results of a pragmatic randomized trial would have been had the trial been conducted in a different population, that is to transport the results from the current target population under study to a new target population. Methods for transporting effect estimates (standardization, inverse weighting, or doubly robust methods) require adjustment for all causes of the outcome that are differently distributed between the trial sample and the new target population (53–57).

We do not make specific recommendations about external validity since this is an evolving area of research. However, it is important that the target population of interest be considered during the design phase of a pragmatic randomized trial and efforts made to ensure that the results of the trial will be applicable to that population.

## 5. Estimation of the per-protocol effect for point interventions

A point intervention is an intervention that takes place only once at the start of follow-up. A sustained intervention or treatment strategy, on the other hand, is sustained over time after baseline. For example, an intervention that consists of a screening test at baseline and lets individuals do whatever they wish after baseline is a point intervention, whereas an intervention that consists of a screening test at baseline and no additional tests during the follow-up is a sustained strategy. Note that treatment assignment is a point intervention (it only occurs once at the time of randomization) and thus the intention-to-treat effect is a contrast of point interventions. This section focuses on the estimation for per-protocol effects for point interventions. The next section focuses on sustained strategies.

Because a point intervention is delivered at or close to the time of randomization, only covariates at or before the time of randomization can influence adherence to a point intervention. To validly estimate the per-protocol effect, baseline variables which predict adherence and are prognostic for the outcome need to be accounted for, either through direct adjustment or via an instrumental variable analysis. Yet two commonly used analytic approaches do not incorporate any such adjustment:

- Naïve per-protocol analysis, that is, restricting the analytic subset to adherent individuals
- As-treated analysis, that is, comparing individuals based on the treatment they choose

Therefore, neither approach can generally provide unbiased estimates of the per-protocol effect when adherence does not occur at random (see Appendix) (12). For example, in a trial of usual care vs. a new treatment, suppose that clinicians often put individuals with more severe disease on the new treatment (e.g., due to a perceived greater need) than in usual care, regardless of their randomized assignment. Then individuals on the new treatment will be, on average, sicker than the ones using usual care, and any analytic approach that directly compares users of new treatment vs. usual care will result in invalid per-protocol effect estimates.

Here we consider two approaches to the estimation of per-protocol effects: direct adjustment for measured confounders and indirect adjustment based on instrumental variable estimation. In what follows we use the term "per-protocol analysis" to refer to analyses whose goal is the estimation of the per-protocol effect. Also, note that valid estimation of the per-protocol effect, like that of the intention-to-treat effect, requires appropriate adjustment for selection bias due to losses to follow-up and a choice of causal estimand when competing events exist. Therefore, the guidelines described above for intention-to-treat effects also apply to per-protocol effects and we will not repeat them here.

*5.1 Direct adjustment for confounders*

One option for validly estimating the per-protocol effect in a pragmatic trial with a point intervention is to directly adjust for baseline prognostic factors that are also predictors of adherence, i.e. baseline confounders. Many statistical approaches are valid to adjust for confounders in per-protocol analyses. Two of them, inverse probability and standardization (58–61), allow calculation of absolute risks in the study population and are therefore preferable. Other commonly used adjustment methods, like outcome regression and propensity score adjustment or matching, typically make strong assumptions about no effect heterogeneity, and do not easily yield unconditional absolute risks.

A good practice for any per-protocol analysis is to include an outcome that can serve as a negative control (62,63). That is, an outcome known to be unaffected by treatment and whose association with treatment would suggest the presence of residual confounding for the primary outcome (see Appendix).

**Guideline:**
IX. **When sufficient data on baseline confounders exist, estimate the per-protocol effect of point interventions via adjustment by inverse probability weighting, standardization, doubly-robust estimation, or other methods.**

*5.2 Using the randomized assignment as an instrumental variable*

When information on baseline confounders is not available, or we do not know why people adhere, sometimes a per-protocol effect for point interventions can be quantified via an instrumental variable analysis with randomization assignment as the instrument (64). The validity of instrumental variable methods requires several conditions, not all of which are guaranteed to hold in randomized trials. Specifically, the randomized assignment needs to meet three conditions to qualify as an instrumental variable (or instrument). An informal description of these instrumental conditions follows.

First, randomization assignment must be predictive of treatment received. This is expected to hold in all trials, and can be empirically confirmed. Second, the effect of randomization assignment must be unconfounded. This cannot be confirmed empirically but is expected to hold in randomized trials. Note, however, that loss to follow-up and competing events can affect the validity of this assumption for instrumental variable analysis and may require covariate adjustment similar to that described in Sections 4.2 and 4.3. Third, the effect of treatment assignment on the outcome must be entirely mediated through the received treatment. This third condition, often referred to as the exclusion restriction, cannot be empirically verified and is not guaranteed to hold by design alone. In particular, this condition is less likely to hold in pragmatic trials that are unblinded (and therefore participants can alter their outcome risk through changes to behaviors or treatments) or which compare active treatments (where some participants do not adhere by taking neither treatment) (65,66).

When the three instrumental conditions are expected to hold, randomized assignment can be used as an instrument to estimate upper and lower bounds for the per-protocol effect (Case Study H). That is, an instrumental variable does not contain enough information to obtain a point estimate of the causal effect, but only estimates of the upper and lower bounds of the causal effect (each bound has its own 95% confidence interval). These bounds are usually too wide and therefore relatively unhelpful to guide decision making, but the bounds can be narrowed down by making additional assumptions (67,68).

*Case study H: Estimating bounds on the per-protocol effect*

> The Norwegian Colorectal Cancer Prevention (NORCCAP) trial was a primary prevention trial for colorectal cancer incidence and mortality comparing a one-time screening sigmoidoscopy with no screening (18). Since this trial assessed a point intervention, bounds for the per-protocol effect could be obtained if the randomized assignment meets the three instrumental conditions.
>
> The first condition holds because individuals randomized to usual care were unable to obtain a screening sigmoidoscopy (not available in Norway) and because 65% of individuals randomized to the screening arm received a screening sigmoidoscopy. The second condition is expected to hold because randomization arm will be unconfounded by design. Finally, we assume that the exclusion restriction holds because it is unlikely that individuals could or would alter their risk of death or colorectal cancer after receiving the invitation letter for a sigmoidoscopy.
>
> In the NORCCAP trial, the age-standardized intention-to-treat difference in 10-year risk of all-cause mortality was -0.2 percentage points (95% CI: -0.6, 0.2). Under the three instrumental variable assumptions, the age-standardized per-protocol difference in the 10-year risk of all-cause mortality was bounded between -5.9 percentage points and 29.3 percentage points (68).

**Guideline:**
X.  **When the three instrumental conditions are expected to hold for treatment assignment, estimate bounds for the per-protocol effect of point interventions. Provide a justification for why you believe the exclusion restriction holds, including performing appropriate falsification tests.**

When the three instrumental conditions *and* another unverifiable fourth condition is met, randomized assignment can be used as an instrument to obtain a point estimate of a per-protocol effect. A variety of fourth conditions exist, including different versions of effect homogeneity and monotonicity. Importantly, the choice of fourth condition determines the type of per-protocol effect that is targeted. Estimation under different forms of effect homogeneity leads to the per-protocol effect that we have considered throughout this section: the average effect in all individuals in the study population. However, effect homogeneity is often perceived as unrealistic because it requires that there is either a constant treatment effect among everyone or, informally, that there are no unmeasured modifiers of the effect of treatment on the outcome (66).

Therefore, monotonicity is (implicitly or explicitly) used in most instrumental variable analyses. Informally, monotonicity means that no individuals in the trial would take exactly the treatment they were not assigned to, regardless of which treatment they were assigned to (64,69). The absence of these "defiers" seems to be a reasonable assumption in many trials, but estimation under monotonicity leads to a different per-protocol effect: the average effect in the subset of individuals who have the unobserved characteristic that they adhered to their assigned intervention and would have also adhered if they had adhered to the other intervention. Individuals in this unidentifiable subset of the population are referred to as "compliers", in the sense that they would have complied with any treatment they would have been assigned to, and the effect in the compliers as the local average treatment effect, or LATE (Case Study I).

Unfortunately, the subset of "compliers" is not usually identifiable. We can never know if a given adherent individual is someone who would have always adhered regardless of which treatment they had been assigned, or whether they only adhered in the actual trial because of the treatment they were in fact assigned. For example, some individuals who adhere to the comparator treatment may not have adhered to the investigational treatment because they would have experienced intolerable side effects on that treatment. Further, identifying those individuals outside of the trial participants for whom the LATE is relevant can be even more challenging.

Therefore, in instrumental variable analyses under monotonicity, a discussion is needed as to why a per-protocol effect in the "compliers" is of interest. Also, the analysis needs to quantify the proportion of "compliers" in the study population and their characteristics compared with the rest of the population (interestingly, both these things can be achieved even though identifying individual "compliers" is not possible) (70).

> *Case study I: Estimating the per-protocol effect in a subset using randomization as an instrument*
>
> In the NORCCAP trial, it is also reasonable to assume monotonicity of the instrument (randomization). In fact, the one-sided non-compliance in the NORCCAP trial guarantees that monotonicity holds – that is, that there are no individuals who would have refused a screening sigmoidoscopy if randomized to that arm, but have somehow obtained one if randomized to usual care. Under this assumption and the three instrumental conditions, we can use randomization as an instrument to estimate the per-protocol effect among the compliers
>
> Among women, the estimated per-protocol difference in 15-year risk of colorectal cancer incidence in the treated was -0.27 percentage points (95% CI: -0.72, 0.18) (18). For comparison, the estimated intention-to-treat difference in 15-year risk of colorectal cancer incidence was -0.19 percentage points (95% CI: -0.49, 0.11) (18).

**Guideline:**
   **XI.  When the three instrumental conditions and monotonicity are expected to hold, discuss whether the effect in the "compliers" is of interest. If so, estimate it and provide information on the relative size and characteristics of the "compliers" subset.**

## 6   Estimation of the per-protocol effect for sustained treatment strategies

Sustained treatment strategies are those that start at baseline and continue during the follow-up. For example, "take daily medication" or "never receive a screening colonoscopy during follow-up". Sustained treatment strategies can be further classified into static strategies, under which all individuals take the same treatment during the follow-up (e.g., "take aspirin every day no matter what happens"), or dynamic strategies, under which the treatment received by each individual depends on his/her evolving characteristics (e.g., "take aspirin every day until a diagnosis of stroke"). Dynamic treatment strategies are usually the most relevant for clinical decision-making.

When estimating the per-protocol effect of sustained treatment strategies the first step is to specify the strategies. Ideally, a precise definition of the strategies would be found in the trial protocol. However, it is not uncommon to encounter situations in which the protocol is ambiguous regarding the strategies being studied. For example, many protocols give study clinicians discretion to advise individuals who experience mild side effects to discontinue treatment. Then the treatment strategy may be a dynamic strategy that allows discontinuation after side effects than a static strategy of continuous use of trial medications. In fact, the dynamic strategy is arguably the most clinically relevant and therefore the one that should be considered when estimating the per-protocol effect (Case study J).

In addition, realistic treatment strategies often need to incorporate a grace period during which treatment decisions, such as initiation, dosage adjustment, discontinuation, or switching, are allowed to occur. For example, consider a protocol that specifies that treatment initiation should occur after reaching a particular threshold (e.g., exceeding a systolic blood pressure of 120 mm Hg). Because it would be unrealistic to expect that treatment initiation will occur exactly on the same day as the threshold is reached, the protocol should also specify a grace period (say, 2 months after the threshold) during which a treatment initiation will be considered adherent. The existence of a grace period means that the treatment strategy is sustained.

> *Case study J: Specifying the protocol when estimating per-protocol effects for safety outcomes*
>
> The Saxagliptin Assessment of Vascular Outcomes Recorded in Patients with Diabetes Mellitus—Thrombolysis in Myocardial Infarction 53 (SAVOR-TIMI 53) trial was a double-blind, placebo-controlled randomized clinical trial designed to assess the safety and efficacy of saxagliptin for cardiovascular outcomes among patients with diabetes mellitus (71). Participants were randomized to initiate either saxagliptin or placebo and continue receiving treatment until symptoms of renal impairment developed, at which point they would have a single dosage adjustment, and no patients were allowed to take other DPP-4 inhibitors or glucagon-like peptide 1 agonists (71).
> One possible interpretation of the trial protocol is that non-adherence includes initiation of any other DPP-4 inhibitor or glucagon-like peptide 1 agonist, or discontinuation of the drug at any time. Another possible interpretation of the trial protocol is that non-adherence also includes failure to decrease study drug dose upon incident renal impairment, or alterations in dosage of study drug other than at the time of incident renal impairment. Each of these definitions would have different implications for data analysis.

**Guideline:**

**XII. To estimate the per-protocol effect of sustained treatment strategies, specify *a priori* a treatment protocol that incorporates real world clinical decision-making, including discontinuation, switching, or dose-reduction rules. When there is sufficient ambiguity about the appropriate treatment strategies, more than one protocol strategy can be specified.**

Because a sustained strategy is delivered during the follow-up, covariates at any time before or after randomization can influence adherence. That is, the estimation of per-protocol effects for sustained interventions require adjustment for pre- and post-randomization prognostic factors that predict adherence during the follow-up.

As with point interventions, there are two basic approaches to estimating per-protocol effects of sustained treatment strategies—direct and indirect adjustment for post-randomization confounding. First, we can directly adjust for pre- and post-randomization confounders using g-methods—inverse probability weighting, the g-formula, or g-estimation of structural nested models—which require the same untestable assumptions that are needed for causal analyses of observational studies. Second, we can use extensions of g-estimation which generalizes instrumental variable estimation to the setting of sustained treatment strategies. This second approach, which requires the instrumental assumptions plus detailed modeling assumptions about the effect of treatment on the outcome of interest (72–74), has been rarely used and will not be discussed here (75–78).

*6.1 Data requirements*

In addition to the information required to estimate the per-protocol effects of point interventions, valid estimation of per-protocol effects of sustained treatment strategies requires accurate information on a variety of post-randomization variables. These include variables that are needed to determine adherence to the treatment strategy (e.g., treatment use and dose, contraindications, adverse effects) and post-randomization confounders.

Trials with sparse data or non-systematic collection after randomization are less likely to be useful for the estimation of per-protocol effects. Simulation studies confirm that the bias for the per-protocol effect increases as the inter-visit interval gets larger (79), whereas the intention-to-treat effect estimate is not affected by the frequency of study visits, since it does not depend on adjustment for post-randomization confounders.

When frequent visits are not possible, linking to electronic health records may be an alternative method of collecting data on post-randomization variables. However, careful thought should be given to the completeness, accuracy, and usefulness of this information source for the purposes of the trial (see Appendix).

**Guideline:**
**XIII.**    **Ensure that sufficient data are collected to determine whether participants adhered to their assigned strategies throughout the follow-up, and to adjust for time-varying prognostic factors that predict adherence to the assigned treatment strategies.**

*6.2 Methods to adjust for post-randomization confounders*

When comparing sustained treatment strategies, the post-baseline treatment is time-varying and therefore the post-baseline confounders are also time-varying. In many clinical settings, the time-varying confounders are themselves affected by prior treatment, and thus we say that there is treatment-confounder feedback. For example, when comparing the effect of epoetin dosing strategies on the mortality of individuals with end-stage renal disease, the hemoglobin value is a time-varying confounder because hemoglobin is a prognostic factor that is used to decide the epoetin dose and, in addition, there is treatment-confounder feedback because prior epoetin doses affect subsequent hemoglobin values.

Conventional adjustment methods cannot handle time-varying confounders when there is treatment-confounder feedback. In fact, conventional methods such as multivariate outcome regression, stratified analyses, propensity score regression and matching, and others will be biased when time-varying confounders are omitted and may introduce bias when time-varying confounders are included in the models (49,80–82).

In contrast, g-methods, developed by Robins and collaborators since 1986, can appropriately adjust for measured time-varying confounders in the presence of treatment-confounder feedback (11,80). The three classes of g-methods are inverse probability weighting (Case Study K), the plug-in g-formula (Case Study L), and g-estimation of structural nested models (Case Study M), as well as their doubly robust versions. The Table in reference (3) provides a comparison of the requirement of each of the g-methods.

Of course, g-methods will result in biased estimates of the per-protocol effect if not all important pre- and post-baseline confounders are available for adjustment.

*Case study K: Estimating the per-protocol effect of sustained strategies using inverse probability weighting*

Participants in the Prevención con Dieta Mediterránea (PREDIMED) trial were randomized to their usual diet with either supplemental olive oil or nuts, or to a low-fat diet (83). The primary outcome was a composite of coronary heart disease events and death from cardiovascular causes over the 6-year follow-up period. The per-protocol analysis used inverse probability weighting to adjust for confounding by baseline and post-randomization variables associated with adherence and prognostic for the outcome. In addition, some participants did not attend all study visits. Therefore, the per-protocol analysis also used inverse probability weights to adjust for selection bias due to loss to follow-up.

The intention-to-treat rate difference was 12.9 fewer cases (95% CI: 5.4, 21.1) of the combined endpoint per 1000 persons after 3 years for Mediterranean vs, low-fat diet. The corresponding per-protocol estimate was 21.3 fewer cases (95% CI: 3.8 to 44.8) per 1000 persons.

The g-formula and inverse probability weighting can be used to obtain estimates of absolute risks and risks differences. Inverse probability weighting is a simpler method to implement but it may result in imprecise estimates when it cannot be combined with a dose-response marginal structural model, which is often the case when comparing dynamic strategies (21). The g-formula is a more flexible approach that results in more precise estimates, but its validity depends on the correct specification of models for all time-varying confounders and the outcome. Theoretically, the g-formula effect estimates will not be null even if the null hypothesis is true. However, in practice, random variability appears to overwhelm any bias due to the g-null paradox.

*Case Study L: Estimating the per-protocol effect of sustained strategies using the parametric g-formula*

The Strategic Timing of AntiRetroviral Treatment (START) trial was designed to assess the impact of antiretroviral treatment initiation strategies on a combined primary outcome of AIDS events or all-cause mortality, among individuals with HIV who were treatment naïve and had good immune function at baseline (CD4 count above 500 cells/ml). Individuals were randomly assigned to either immediate treatment initiation at enrollment or delayed initiation when immune function decreased or upon diagnosis of AIDS.

After specifying the protocol, including a grace period for treatment initiation, the per-protocol effect was estimated using the parametric g-formula. The estimated intention-to-treat difference in 5-year risk of AIDS or death was -3.1 percentage points (95% CI: -5.2, -0.8). The corresponding per-protocol difference was -3.8 percentage points (95% CI: -6.7, -1.5) (84).

G-estimation of structural nested models is a more limited approach for the estimation of per-protocol effects because it does not readily yield estimates of absolute risk and it cannot easily accommodate complex dynamic strategies. It also needs to be combine with inverse probability weighting when adjustment for selection bias due to loss to follow-up is necessary (50,85).

> *Case Study M: Estimating the per-protocol effect of sustained strategies using g-estimation of structural nested models*
>
> In Case Study #6, we described a randomized trial of antipsychotic use in which individuals with schizophrenia were randomized to either atypical or conventional antipsychotics. After adjustment or loss to follow-up via inverse probability weighting, the estimated intention-to-treat difference in BPRS score was -0.86-units (95% CI: -3.88, 2.15) for atypical vs. conventional antipsychotics. The estimated per-protocol difference in BPRS score was -1.50-units (95% CI: -6.84, 3.84) after adding confounding adjustment via g-estimation of a structural nested model (50).

**Guideline:**
**XIV.    Use g-methods to adjust for time-varying confounders when there is treatment-confounder feedback. Choose inverse probability weighting or the g-formula to obtain adjusted estimates of absolute risks and risk differences.**

## 7    Discussion

Pragmatic randomized trials are a useful tool for estimating the comparative effectiveness of treatment options. However, the features of these trials which are most useful for clinical decision-making make estimation more challenging by introducing the possibility of confounding, selection bias, and competing risks. Until now, there have not been any clear guidelines for the valid estimation of causal effects from pragmatic trials.

We propose fourteen guidelines for the analysis of pragmatic randomized trials to address this gap (Table 1). These guidelines fall under four broad categories:
- Choice of causal effect
- Estimating the intention-to-treat effect (these guidelines are also relevant to the per-protocol effect)
- Estimating the per-protocol effect of point interventions
- Estimating the per-protocol effect of sustained interventions

Since the intention-to-treat effect will often provide limited information for clinical decision-making, we strongly urge that all pragmatic randomized trials should be designed to allow for valid estimation of both the intention-to-treat and the per-protocol effect.

A number of additional concerns may arise in the design and analysis of pragmatic trials for which insufficient theoretical basis is available to provide clear guidelines. In particular, there are currently no simple methods for estimating the required sample size under an expected degree of non-adherence or loss to follow-up when using g-methods. Similarly, more work is needed to compare the available methods for g-estimation when sustained treatment strategies are of interest and minimal covariate information is available.

Finally, *a priori* specification of the statistical analysis plan is an important tool for ensuring replicability and preventing p-hacking or 'hypothesizing after results are known' (HARK-ing) (86). However, for per-protocol effect estimation, full *a priori* specification of the statistical analysis plan is often difficult since the degree of non-adherence or loss to follow-up, and reasons for these post-randomization events, that will be observed in the trial is unknown before the trial begins. The statistical analysis plan may therefore need to include adaptive features, such as rules for modeling inverse probability weights, sensitivity analyses for the assumptions required in per-protocol effect estimation, or the use of alternative methods which make use of different assumptions (14). Further work is needed to clarify best practices for these adaptive statistical analysis plans in pragmatic randomized trials.

Table 1. Guidelines for the estimation of causal effects from pragmatic trials

| Category | Guideline |
|---|---|
| Choice of causal effect | I. To adequately guide decision making by all stakeholders, report estimates of both the intention-to-treat effect and the per-protocol effect, as well as methods and key conditions underlying the estimation procedures. |
| | II. Report absolute risks and their differences, as well as their ratios, for discrete outcomes. |
| | III. Heterogeneity of treatment effects can be reported using subgroup analyses that use the additive scale to measure the effect of interest. Patients and advocates should be included in *a priori* specification of subgroups. |
| Estimating the intention-to-treat effect* | IV. Pre-specify important prognostic factors for the outcome and the maximum acceptable difference in the distribution of these factors between groups. When one or more prognostic factor meets the threshold for imbalance, adjust via standardization, inverse probability weighting or, preferably, doubly-robust methods. |
| | V. In sensitivity analyses, adjust for large imbalances in any important prognostic factors, regardless of whether they have been pre-specified. |
| | VI. In survival analyses with competing events, report both the risk of the competing event and the risk of the event of interest among those who survived the competing event by treatment group. |
| | VII. In survival analyses with competing events, specify the intention-to-treat effect as the total effect of treatment assignment on the outcome of interest (the simplest analysis), and justify interest in any additional effects that are estimated. |
| | VIII. Ensure that the trial protocol specifies the collection of post-randomization time-varying prognostic factors that predict loss to follow-up, and appropriately adjust for these factors to reduce selection bias. |
| Estimating the per-protocol effect of point interventions | IX. When sufficient data on baseline confounders exist, estimate the per-protocol effect of point interventions via adjustment by inverse probability weighting, standardization, doubly-robust estimation, or other methods. |
| | X. When the three instrumental conditions are expected to hold for treatment assignment, estimate bounds for the per-protocol effect of point interventions. Provide a justification for why you believe the |
|  | exclusion restriction holds, including performing appropriate falsification tests. |
|  | XI. When the three instrumental conditions and monotonicity are expected to hold, discuss whether the effect in the "compliers" is of interest. If so, estimate it and provide information on the relative size and characteristics of the "compliers" subset. |
| Estimating the per-protocol effect of sustained interventions | XII. To estimate the per-protocol effect of sustained treatment strategies, specify a priori a treatment protocol that incorporates real world clinical decision-making, including discontinuation, switching, or dose-reduction rules. When there is sufficient ambiguity about the appropriate treatment strategies, more than one protocol strategy can be specified. |
|  | XIII. Ensure that sufficient data are collected to determine whether participants adhered to their assigned strategies throughout the follow-up, and to adjust for time-varying prognostic factors that predict adherence to the assigned treatment strategies. |
|  | XIV. Use g-methods to adjust for time-varying confounders when there is treatment-confounder feedback. Choose inverse probability weighting or the g-formula to obtain adjusted estimates of absolute risks and risk differences. |

\* These guidelines are relevant to both the intention-to-treat and per-protocol effects

**Appendix**

*A.1. Methods that are generally not valid for estimating the per-protocol effect*

A variety of other per-protocol analysis methods exist but generally do not provide valid estimates of the per-protocol effect. In addition, the terminology used to describe these methods can introduce unnecessary complexity and confusion because the target causal effect (or estimand) is generally the same regardless of analytic approach. We briefly summarize these approaches and describe why they do not provide valid estimates of the target estimand.

A naïve *per-protocol analysis*, also called an *on-treatment analysis*, is one in which non-adherent person-time is censored from the analytic dataset and no adjustment is made (6). An *as-treated analysis* is but allows individuals to switch (or cross-over) between treatment arms without censoring, and individuals or person-time are analyzed based on treatment received, rather than by randomization arm. These analyses are both attempts to estimate the effect of continuous receipt of the investigational treatment versus the comparator treatment on the outcome, but failure to adjust for predictors of treatment received leads to bias whenever these variables are also prognostic for the outcome (12). Further, the continuous treatment effect is not necessarily the per-protocol effect of interest.

Finally, a *modified intention-to-treat analysis* is generally a form of per-protocol effect estimation rather than intention-to-treat effect estimation. In this analysis, any individual who does not initiate at least one dose of assigned treatment is removed from the analytic dataset. The target estimand for a point intervention is the effect of adherence to the investigational treatment versus adherence to the comparator treatment on the outcome, and for a sustained intervention is the effect of adherence to initiation of the investigational treatment versus adherence to initiation of the comparator treatment. In general, no adjustment is made for confounders of treatment initiation leading to a biased estimate of this causal estimand.

These conventional methods for per-protocol effect estimation generally do not provide valid estimates of their target estimands because they are commonly implemented without adjustment for predictors of adherence, switching, or initiation that are also prognostic for the outcome. Unless these confounders are appropriately measured and adjusted for, the estimates obtained via these methods will therefore be biased. Only when treatment decisions are made completely at random are these four approaches valid; these methods then become special cases of the per-protocol effect estimation methods described in the main text.

*A.2. Effects under other protocols*

Many trials may also be interested in estimating causal effects which deviate from the randomized interventions. We briefly discuss three examples of these effects.

In many oncology trials, there are ethical concerns about withholding effective treatments from patients randomized to the comparator arm. These trials may be designed with a protocol that allows participants randomized to the comparator arm to switch to the investigational treatment after some pre-specified time period, such as upon disease progression. In these trials, the per-protocol effect would consider individuals who switch as adherent. In some cases, investigators may still be interested in knowing the effect had no one switched, which we could call the switching-free effect (87). This switching-free effect requires many of the same assumptions as the per-protocol effect, but also requires information on time of switching, and on confounders for switching and the outcome. However, it should be noted that the switching-free effect may not be clinically relevant since it may be unlikely that oncologists would ever retain patients on their primary treatment after disease progression.

The (controlled) direct effect of treatment is the effect to receiving the intervention that is not mediated through a specific third variable (a mediator). For example, the initial results of Women's Health Initiative (WHI) trial were surprising to many. The estimated intention-to-treat effect on coronary heart disease incidence if all women had been assigned to hormone therapy versus if they had been assigned to placebo was a hazard ratio of 1.24 (95% CI: 1.00, 1.54) (29). However, women in the hormone therapy arm were less likely than women in the placebo arm to initiate statins during follow-up. Therefore, a secondary analysis of the trial data attempted to assess whether the apparent effect of hormone therapy on coronary heart disease could be explained via statin initiation, using a structural nested model approach to estimate the controlled direct effect of hormone therapy if everyone had initiated statins (88). Interestingly, the controlled direct effect did not differ meaningfully from the main trial results, and so it appeared that trial findings were unlikely to be explained through differences in statin use during follow-up (88).

Another scenario where the controlled direct effect of treatment is of interest is when changes to the standard of care occurs during the trial follow-up period (5). For example, in the ACTG 002 (AIDS Clinical Trials Group 002) trial, participants with HIV were randomized to the use of high- versus low-dose zidovudine, and no guidance was made about the use of prophylaxis for opportunistic infections. During the trial, the use of prophylaxis for *Pneumocystis* pneumonia differed between trial arms, but by the end of the trial the use of prophylaxis had become standard practice. Therefore, a relevant clinical question was whether the low-dose group (which had better survival in the intention-to-treat effect) would have still had better survival than the high-dose group had everyone in the trial received prophylaxis. Analysis of this direct effect requires the use of g-methods(5).

Finally, investigators may be interested in estimating the effects of other, non-randomized treatments within the trial population. In this case, the trial population can be considered an observational cohort and adjustment for baseline and post-baseline confounders using g-methods is required (89).

*A.3. Missing adherence data*

Unlike the intention-to-treat effect where treatment assignment is almost always known for all individuals enrolled in the trial, per-protocol effects rely on information about exposure status collected over time after randomization. It is not only possible but common that exposure information will be missing for at least some visits for at least some individuals. In order to estimate the per-protocol effect including these individuals some decision must be made about the missing exposure data (Case Study N). A common approach is to either censor these individuals without adjustment or to assign them an exposure value often based on single-value imputation (for example, to assume that all missing adherence implies non-adherence). Both of these approaches can lead to bias whenever adherence information is missing conditionally at random or missing not at random. However, if sufficient baseline and post-randomization data on predictors for adherence measurement exist, inverse probability of measurement weights can be used to up-weight those individuals or time points with complete adherence data.

| *Case study N: adjusting for missing adherence data when estimating the per-protocol effect* |
|---|
| The Coronary Drug Project (CDP) trial compared the mortality of adherers and non-adherers among those assigned to the placebo arm. In order to maximize available data, they categorized any person-time with missing adherence information as non-adherent. This made the strong assumptions that |

> missingness was highly informative of adherence value, but not associated with any measured or unmeasured covariates. These assumptions were likely not correct and introduced potential bias into their analysis, because missingness was also highly predictive of mortality (90). In a re-analysis of this trial, we instead censored individuals when their adherence status was unknown for more than a year and used inverse probability of censoring weights to adjust for loss to follow-up by individuals who stopped reporting adherence information. This removed the bias caused by inappropriate imputation of missing adherence (91,92). The unadjusted 5-year mortality risk difference comparing placebo adherence to non-adherence when missing data was assumed to be non-adherence was 14.3 percentage points (95% CI: 10.8, 17.8), whereas the unadjusted risk difference was 11.0 percentage points (95% CI: 6.5, 15.6) when missing data was carried forward for intermittent missingness and censored after 3 consecutive missed visits. (Note that improving the assumptions about missing adherence data was not sufficient to fully address the bias in the unadjusted analysis – adjustment for baseline and post-randomization confounders was still required.)

*A.4. Sensitivity analyses and falsification tests*

The recommended analytic methods for estimating intention-to-treat and per-protocol effects under loss to follow-up, competing events, and non-adherence discussed in these guidelines require strong assumptions more commonly applied to observational studies. Few of these studies can be empirically verified, and good statistical practice therefore requires the use and reporting of sensitivity analyses to determine the impact of potential violations of these assumptions on the estimated effects.

For point interventions, sensitivity analyses for unmeasured confounding, such as the e-value, can be useful tools for assessing the potential for residual bias (33). When using inverse probability weighting (for loss to follow-up or non-adherence) or propensity score adjustment, appropriate sensitivity analyses will include a description of the distribution of the weights or propensity scores as well as a range of models testing alternative modeling assumptions. Graphical assessment of balance in weights or propensity scores can be useful but does not guarantee unconfoundedness. When using the parametric g-formula, sensitivity analyses should include comparing the observed covariate distribution to the distribution simulated under the treatment assignment probability, as well as assessment of the sensitivity of results to model specification assumptions. For all confounding-adjustment based methods, negative control outcomes can be a useful tool for assessing the potential for residual confounding. In trials with a placebo arm, adherence to placebo versus non-adherence to placebo can be used as a negative control exposure to partially assess the assumptions required for estimating the per-protocol effect(23,91). Finally, for instrumental variable methods, falsification tests should be performed to assess the potential for residual bias in the estimate (93,94).